# The Complexity of Human Computation: A Concrete Model with an Application to Passwords

Manuel Blum[1] and Santosh Vempala[2]

4 July 2017

**Abstract**. What can humans compute in their heads? We are thinking of a variety of Crypto Protocols, games like Sudoku, Crossword Puzzles, Speed Chess, and so on. The intent of this paper is to apply the ideas and methods of theoretical computer science to better understand what humans can compute in their heads. For example, can a person compute a function in their head so that an eavesdropper with a powerful computer – who sees the responses to random inputs – still cannot infer responses to new inputs? To address such questions, we propose a rigorous model of human computation and associated measures of complexity. We apply the model and measures first and foremost to the problem of (1) humanly computable password generation, and then consider related problems of (2) humanly computable "one-way functions" and (3) humanly computable "pseudo-random generators".

The theory of Human Computability developed here plays by different rules than standard computability, and this takes some getting used to. For reasons to be made clear, the polynomial versus exponential time divide of modern computability theory is irrelevant to human computation. In human computability, the step-counts for both humans and computers must be more concrete. Specifically, we restrict supercomputer adversaries to at most $10^{24}$ (Avogadro's number) steps[3,4].

An alternate view of this work is that it deals with the analysis of algorithms and counting steps for the case that inputs are small as opposed to the usual case of inputs large-in-the-limit.

Application and running example: password generation schemas are humanly computable algorithms based on private keys. Humanly computable and/or humanly usable mean, roughly speaking, that any human needing – and capable of using – passwords can if sufficiently motivated generate and memorize a secret key in less than one hour (including all rehearsals), and can subsequently use schema + key to transform website names (challenges) into passwords (responses) in less than one minute. Moreover, the schemas have precisely defined measures of security against all adversaries, human and/or machine.

## 1. Introduction

The processing power of humans is much more limited than that of computers for simple grade school arithmetic. Though humans can do some amazing highly skilled mental jobs that computers are only now beginning to do, like write gripping stories, state hard mathematically interesting problems, discover laws of nature, etc., they (humans) are limited in their ability to do arithmetic computations in their heads[5]. In particular, humans have only a tiny short-term memory for performing arithmetic operations. They also do arithmetic operations much more slowly than computers. Modern complexity theory and cryptography are designed for computers (Turing machines) and/or for humans with access to computers, but not for humans working alone in their head. This paper is concerned with investigating problems that might possibly be solved by humans working alone - using humanly usable algorithms (single agent) or humanly usable protocols (multiple agents). The broader problem considered here is

---

[1] Carnegie Mellon University, mblum@cs.cmu.edu
[2] Georgia Tech, vempala@gatech.edu
[3] Why this particular number? The intent here is to have as example a SPECIFIC MEMORABLE number (in this case, Avogadro's number) that is roughly the minimum size sufficient to keep adversaries at bay – adversaries that have a week's time on the fastest 2016-17 supercomputer, which in January 2017 can do $10^{17}$ petaflops/sec = $6x10^{22}$ flops/week. In general, the specific number will depend on current technology.
[4] While we use specific values and give justifications for them everywhere, most, like Avogadro's number, are actually placeholders for more general constants. In general, a cryptographic schema's analysis will have bounds on usability and security. Usability bounds are upper bounds on the maximum amount of work a human should be expected to do. Security bounds are lower bounds on the minimum amount of work an adversary must do to "break" the schema.
[5] Human-oriented schemas can ask humans to use their (currently) distinctly human computing powers to generate passwords. For example, a human might be shown a photograph of a street scene and asked to count the number of humans, dogs, or bicycles in the scene. In this work, we look only at schemas that transform website names, i.e. character strings not photographs, into passwords. See [Blocki 2014] for schemas that accept pictorial challenges.

THE PROBLEM:

> What functions can a human compute in his/her head that a powerful human-computer adversary cannot break[6] from observing a specifically limited amount of input-output behavior? Can a human transform (truly) random seeds into "pseudorandom sequences" in his or her head such that the "pseudorandom sequences" are indistinguishable from truly random sequences to a (powerful) PC that is run for at most $10^{24}$ steps? In this work, the adversary is presumed to know the publically available schema, which is the humanly computable algorithm minus the private key, and to observe – possibly participate in – the public communications called for by the protocol.

THE MODEL (Complexity in the Small vs Asymptotic Complexity):

- Since humans are slow at arithmetic computations, at least in comparison to computers, complexity bounds described by functions such as polynomials or exponentials are not suitable for analyzing human computation, and consequently, as we shall see, not appropriate for analyzing Turing Machine adversaries either. We must give up thinking that an exponential function like $10^{x/10}$ is worse than a linear function like $10(x+10)$. Indeed, the entire range of x for human computation may be $0 \leq x \leq 25$, and in that range the exponential $10^{x/10}$ is smaller than the linear $10(x+10)$. This is not a contrived example either: it is easier for a human to decide if 91 is prime using an exponential time factoring algorithm than a polynomial time primality test.

  So instead of demanding that a schema run in polynomial or linear time, we require an exact value for the number of steps that the schema takes ± a small additive constant.[7]

- Our insistence on dealing with **explicit numbers** instead of with **formulas that are correct in the limit** runs completely counter to what is done in complexity theory, but for human computation, this insistence on concrete specific constants rather than asymptotically correct formulas is the right approach. In part, this is because - while computers acquire more memory and run faster with each passing year - human brains remain the same old model (the brain still advances, but through improvements in culture).

- In modern complexity theory, there is a sharp divide between poly-time and exponential-time computability. Cryptographic protocols typically assume that both users and adversaries are randomizing poly-time algorithms. Users can encrypt things in poly time that cannot be decrypted by adversaries that run in poly time. In human computation as defined here, the user is a human working without a computer for a specific bounded amount of time measured in hours, minutes, and seconds. The adversary is a human, similarly bounded, having access to a (two-tape 52-character) Turing machine that can run for at most $10^{24}$ steps.

The following three tasks will serve as motivation and test cases as we set out to address the above question.

1. Humanly usable password schemas
2. Humanly usable "one-way functions"
3. Humanly usable "pseudo-random generators"

The quotes indicate that these "one-way functions" and "pseudo-random generators" are not in general true one-way functions or pseudo-random generators in the sense defined in modern cryptography, which relies on the polynomial versus exponential divide.

Our first task is to come up with a precise definition for humanly usable algorithm (henceforth called a *schema*) that can be used for example to transform challenges (website names) into responses (passwords), all within the acceptable time limits mentioned above, while remaining secure to a well-defined extent from a computationally all-powerful adversary (a Turing machine with unbounded computation time)[8]. Two other tasks are proposals for humanly computable "one-way functions" and humanly computable "pseudo-random generators". The former are functions that humans can compute in their heads, but that an adversary having a PC of clearly specified power (limited to at most $10^{24}$ steps) almost certainly cannot invert.

---

[6] Break = infer, invert, … depends on context.
[7] Even to expect that the composition $f(g(\Box))$ of two humanly computable functions $f(\Box)$ and $g(\Box)$ will be humanly computable is wrong. It may be that f and g are humanly computable but that their composition is not, as occurs for example when the human is not able to store the intermediate output of g in its limited short-term memory.
[8] It would suffice to consider a Turing machine that may run for no more than $10^{24}$ steps.

To state and verify humanly usable schemas and protocols, we first need a definition of humanly usable. For this we define a formal model of human computation and propose two complexity measures: 1) the human cost of preprocessing (PREP), which is about memorizing the schema and generating and memorizing a private key, if any, and 2) the human cost of processing (PROC), which is about using the schema and key to do the required task, our first such task being to transform challenges into responses.

In the rest of this introduction, we discuss the three tasks in more detail. In the next section, we present the model of human computation and the associated costs, with a few examples.

**Password generation.** Passwords are responses to challenges (typically website names). In this paper, passwords are produced by password schemas, which are humanly computable algorithms for mapping (challenge, key) pairs to passwords. The insecurity of commonly used passwords (see e.g., [Ives et al., 2004; Bonneau, 2012; Li et al., 2014]) and the difficulty of memorizing multiple long passwords [Kruger et al., 2004; Shay et al., 2014; Blocki, 2014] have been discussed extensively in the literature. Passwords should be easy to produce when needed, and hard for an adversary to forge even if the adversary knows the user's schema and has seen the passwords to a small number of websites. More generally, we seek schemas that are Analyzable, Publishable, Humanly Usable, Secure and Self-Rehearsing. Analyzable means that the schema is so precisely defined that a Turing machine can execute it. Publishable means that the schema itself (but not the user's private key) is or can be made public. Humanly Usable means three things: 1) the schema itself (but not the key) must be learnable in a few minutes, 2) generation and memorization of a key should take at most an hour, preferably no more than 30 minutes, of the user's lifetime, and 3) generating or regenerating a password should take no more than 1 minute, preferably 20 seconds. Regeneration is especially important because passwords in our view should never be memorized but should be regenerated when needed.

A number **Q** specifies security: a password schema is said to have security Q if an adversary who has seen responses to less than Q challenges – each challenge drawn at random with replacement from a well-defined sample space (a dictionary of words with associated probabilities) – has probability less than 1/10 to guess the correct response to a new challenge randomly drawn with replacement[9]. "Self-Rehearsing" means that, in the process of responding to occasional random challenges, the user rehearses every aspect of the schema and key. In a recent paper, [Blum & Vempala, 2015] established the existence of such schemas. Here we improve on them both in terms of theoretical analysis and practical usability.[10]

**One-way functions.** A one-way function (**OWF**) is a function that is efficiently computable, yet not invertible by a poly-time Turing machine [Goldreich, 2004]. By comparison, a humanly computable "one-way function" (**HU–"OWF"**) is a humanly computable function that cannot be inverted in less than $10^{24}$ steps. We present a candidate for a HU–"OWF".

**Pseudo-random versus (note the quotes) "Pseudo-random" generators**. A standard (cryptographically secure) pseudo-random generator (PRG) is an algorithm that takes as input a random string (of digits or characters) of length n and outputs a string of length 2n [11] that is "virtually"[12] indistinguishable from a (truly) random string of length 2n by a poly-time Turing machine [Blum and Micali, 1982; Yao, 1982; Goldreich, 2004]. By comparison, a "pseudo-random generator" ("PRG") is a schema that transforms n-digit input strings into 2n-digit[13] output strings that for n ≥ 20 are "virtually" indistinguishable from random 2n-digit strings by a computer that takes no more than $10^{12}$ steps.[14] A HU-"PRG", while reminiscent of a PRG, need not be a PRG. We present a candidate for a HU-"PRG".

---

[9] Challenges randomly drawn *with replacement* from a dictionary with just one word (probability 1) have Q = 1, i.e. the adversary need see only one challenge response pair to determine the correct response to the next challenge with probability > 1/10.
[10] An example of Q: Many people have a smallish number of passwords for all logins. Typically k < 10 passwords. For the user who has a method – any method - to assign k passwords to all challenges, the adversary can expect to guess the correct password on or before the √k challenge (by the birthday paradox). This implies that for k ≤ 9, Q ≤ 3.
[11] This 2n can be replaced by n+1, which can be extended to 2n.
[12] "Virtually" needs to be made precise.
[13] The 2n cannot be replaced by n+1, since humanly oriented algorithms are not in general composable.
[14] Why this particular number? There are $10^{2n}$ random strings of length 2n, but only $10^n$ challenges of length n for generating $10^n$ pseudorandom strings of this length 2n. Since the adversary knows the schema but not the challenge, she need try no more than $10^n$ challenges to prove that a given random string is not pseudorandom. To make this hard for her to do, we require that $10^n > 10^{12}$. $10^{15}$ would be better than $10^{12}$, but the proof in this footnote does not allow us to claim $10^{15}$.

## 2. Human Computation

For many cryptographic problems and games[15] (e.g., speed chess, Sudoku), human computation consists of a preprocessing phase **PREP** (memorization of a public i.e. published or at least publishable schema plus generation and memorization of a private information or key), and a processing phase **PROC** (a run of the schema with its associated key on some input). We view PROC as the speedy computation (and output) of a function on a given input. This captures our motivating scenario of responding to each challenge with its associated password.

We wish to model humans for games and cryptographic problems, but begin here with the more specialized model of humans for executing password schemas. For this purpose (of executing password schemas), we model a human as follows:

1. The human is a kind of Turing machine, one in which the usual tape is replaced by two random-access memories, one long-term and the other short-term.

2. Long-term memory is potentially infinite – upper bounded only by the usable lifetime of the human and the time it takes to store information in that memory. Storing information (such as schema and key) in long-term memory is slow; reading information from long-term memory, given a pointer to its location in memory, is relatively fast. Storage in long-term memory is permanent provided it is rehearsed on the doubling schedule described by Woźniak and Gorzelańczyk[16].

    Long-term memory is both written to and read from in the preprocessing phase. In the processing phase, long-term memory is used only for reading.

3. Short-term read-write memory is fast but tiny, typically storing 2 or 3 chunks[17], each chunk being a pointer to some item such as a digit or number, a character or word, an image or music clip. In our model, unlike anything we find in the psychological literature, a chunk is a well-defined object, a pointer into memory.

4. In the context of passwords, the input (called the challenge) is presented as a singly linked list with a pointer in short term memory to its leftmost start location. Whatever its location in the challenge, the pointer can be shifted one link right (but not left), or read (past tense), or reset to the start location, in one step.

**Schemas** in general, not just password schemas, are algorithms intended for humans. They are used in association with information stored in permanent memory, which for passwords includes a parameter called the key. Humans may use dice, paper, pencil, and other such tools to generate and memorize private keys and to memorize a public schema (PREP). They must thereafter execute the schema (PROC) in their heads i.e. without using any tools outside of their head. This model will be sufficiently powerful for our password schemas. Towards the end of the paper, we briefly discuss extensions of the model.

A schema-key combination is considered to be **(COMM TIME, PREP TIME, PROC TIME)-humanly usable** if and only if it satisfies the following requirements:

1. COMM TIME is an upper bound on the time to learn the schema. The latter time includes the time to transform a few sample challenges into passwords (from a description of the humanly usable instructions), and all rehearsal time (for the life of the human). COMM TIME (including rehearsals) must be at most 10 minutes.
2. PREP TIME is an upper bound on the time to generate and memorize the schema's associated private key, and all rehearsal time needed to maintain that memory. For passwords, this PREP TIME is required to be at most 2 hours.

---

[15] In this paper, context determines whether "game" denotes an entire game in the usual sense, or a single-move in such a game. Examples of such single-move games include:
  1. Speed chess in which the challenge or input is a chessboard position together with a color, black or white, and the response or output is that color's move.
  2. Sudoku in which the challenge is a typical Sudoku board with digits in certain locations, and the response is placement of one more digit in some location.
  3. Crossword Puzzles in which the input is a crossword puzzle, including its clues, partially completed, and the output is a word inserted to the puzzle.

[16] Woźniak and Gorzelańczyk [1994] have shown that for an item to remain in long-term memory, the item must be rehearsed on a certain *doubling schedule*: Let t denote the time (after the initial memorization) between two successive successful rehearsals. Then the item will remain in long-term memory provided it is rehearsed at times 2t, 4t, 8t, … as measured from the time of completion of the second rehearsal.

3. PROC TIME is an upper bound on the time to run the schema on a single input. For passwords, the PROC TIME is required to be at most 1 minute.
4. A schema uses at most three, preferably at most two pointers (chunks) into long and short-term memory (see model of a human part 3 above).[18]
5. For passwords, the schema and its associated key (both of which are stored in long term memory) are completely self-rehearsing in that each and every instruction is run (this includes following each and every flowchart arrow), and all elements of the key are rehearsed, in a "significant fraction" of challenge-response computations.[19]

From now on in this paper, COMM is considered to be a part of PREP, and COMM TIME is included in PREP TIME.

In general, not just in this paper, PREP TIME and PROC TIME must be specified to nail down human usability. In this paper, unless we say otherwise, we require that for passwords, PREP TIME ≤ 2 hours and PROC TIME ≤ 1 minute.

From here on, schema will refer to password schemas rather than any more general class of schemas.

## Intermission

We now define the two complexity measures (1) **PREP** = **{(**Human Complexity of creating and memorizing a key**)** plus **(**cost of communicating and memorizing a Schema (**COMM**)**)}**, and (2) **PROC** = Human Complexity of Processing. We include COMM within PREP as it is clearly part of the preparation necessary to use a schema; we separate the two aspects of PREP (namely 1.memorizing and understanding the schema, and 2.memorizing the key) as the two appear to be incomparable. Taking PREP and PROC in reverse order:

> ***PROC*** *(Processing Complexity) = total number of reads from long term (permanent) memory plus total number of reads and writes to short term memory while processing a single input.*

To illustrate the measure, we examine the human complexity measure PROC of some natural operations:

1. Set a pointer to an item that is already in long-term memory (e.g. a sentence or telephone number) referred to in the schema or key: Cost=1.

2. Move a pointer into the singly-linked list for a challenge, sentence, telephone number, etc. to the right by 1, or set a pointer to the start or start+1, or to the end or end-1: Cost=1.

3. Operations of +, and × (mod 2,3,4,5,9,10, 11) or $=^?$ on two single digit operands: Cost = number of digits including the logical 0,1 symbols created during the operation. Examples: $4 =^? 3$ and $4+3$ (mod 10) have cost 1. $4+9$ (mod 10), assuming that the addition consists of doing first $4+9 = 13$ and then $13 \bmod 10 = 3$, has cost 2.

4. Apply a map, typically a map from letters to digits, that has been memorized as a hash function: Cost=1.

We note that PROC is a measure of effort similar in certain ways to asymptotic complexity. Just as the actual running time of an algorithm can differ from computer to computer, and even on the same computer depending on its load, the time taken for human computation can differ from one human to another and even for the same human. Unlike asymptotic complexity of algorithms running on Turing machines or RAM computers, the PROC must not be expressed with a leading big-O term: an effective analysis must compute PROC to within a small additive constant. In later sections, we will perform such analyses for select schemas.

Next, we turn to the human complexity measure, **PREP**, of generating and memorizing a key. For PREP, unlike PROC, the human may use a random number generator, paper and pencil to create a random key, and then store it in permanent (human) long-term memory. Once keys are generated and stored in permanent

---
[17] Miller, G.A. (1956), *The Magical Number Seven, Plus or Minus Two: Some Limits on our Capacity for Processing Information,* Psychological Review, 63, 81-97.
[18] While human short-term memory is said to be able to store 7 ±2 chunks, two or three is all that we have found to be humanly usable for PROC. Perhaps some of the other chunks are needed for other functions? We in any case assume that at most two or three chunks will be available for PROC.
[19] "Significant fraction" is defined so that once the total permissible COMM+PREP time has been exhausted, no additional purposeful rehearsal is needed, as the natural computation of responses to random challenges will suffice to ensure that schema and key remain in permanent memory. Its definition assumes that the user will respond to random challenges at a well-defined rate.

memory – and provided the Woźniak and Gorzelańczyk doubling rehearsal schedule [Pimsleur, 1967; Woźniak & Gorzelańczyk, 1994] is followed to keep the key in permanent memory – the random number generator, paper and pencil should no longer be needed.

**PREP** *(Preprocessing Complexity) = (number of tosses of a k-sided die) × ($\log_2 k$) + (number of chunks written to permanent memory (presumably the key and the schema)).*

One measure of PREP complexity is by comparison to commonly memorized quantities. Some examples with rough estimates of their costs are given below. These costs are for long-term <u>permanent memorizations</u> as opposed to the short-term <u>temporary memorizations</u> counted in PROC.

1. Linked list memorization of a 10-digit string of chunks is equivalent to memorizing a random 10-digit phone number. Cost = number of chunks in the string (= 10 for the 10 digit phone number). **Aside**: Once a string is memorized, it itself becomes a chunk.

2. Linked list memorization of a long string of chunks (letters), roughly equivalent to memorizing the alphabet, is something we can do by age 4 or 5. Cost = number of chunks in the string (= 26 for the alphabet).

In both the above cases, people memorize a linked list with a pointer to the start of the list and, in the case of a longer list like the alphabet, several additional pointers into the list. Access to this linked list does not enable us to recite the list quickly from memory backwards (Z Y X … A) – though of course we can learn to do that if we want. The cost of a linked list memorization of a <u>string</u> of chunks, as in 1 and 2 above, is a "fraction" of the cost of a random access memorization of a <u>map</u> from chunks to chunks, as in 3 below. Because these two kinds of memorization are (so) different, we specify for each memorization whether it is a linked list or a random access hash. **Aside:** To memorize a small amount of data such as a telephone number as a hash, we suggest memorizing that data initially as a singly linked list (e.g. as a telephone number $t_1 t_2$…); then using that memory to enable memorizing the same data as a hash function, e.g. 1 ➔ $t_1$, 2 ➔ $t_2$, …

When it comes to preprocessing, humans more easily store information in a linked list than in a (random access) map. When it comes to processing, however, maps are typically more useful than linked lists.[20]

3. Assume that Alfa, Bravo, Charlie… are chunks. Linked list memorization of the list Alfa ➔ Bravo ➔ Charlie…. ➔ Zulu: Cost = 26.

   Random access memorization of the map A ➔ Alfa, B ➔ Bravo, C ➔ Charlie, … Z ➔ Zulu, which is something that Ham Radio Operators, Boy Scouts, and the military do. Cost = number of chunks in the map (2 × 26 = 52 for the Ham Radio map). This is twice the cost of the linked list memorization.

   Random access memorization of the Morse code map: A ➔ •—, B ➔ —•••, C ➔ —•—•, … , which typically takes days if not weeks. Cost = number of chunks in the map. In Morse code, assuming that the chunks are •, —, A, B, … Z, map C ➔ —•—• costs 5, while E ➔ • costs 2. Cost of random access memorization of the 26 letter Morse code is therefore = 108.

4. Linked list memorization of a poem, the lyrics to a song, or the (meaningful) 272-word Gettysburg Address is the kind of thing that middle school students regularly do. Note that the rhyme in poetry, the melody in song, and the meaningfulness of prose help greatly with memorization (none of which is accounted for in our cost measure). Cost = number of chunks, which in the case of the Gettysburg Address is the number of words = 272.

5. Linked list memorization of a <u>random</u> list of chunks, such as the digits of **π**. Most school children learn a few digits of **π**, and a very few learn maybe 20 digits.[21] Cost = number of chunks (which, except in rare cases, are digits).

**PREP** includes **COMM**, the cost of communicating an algorithm to a human. In standard algorithm theory, **COMM** might simply be the length of the program being communicated. In the study of humanly usable algorithms, we propose:

---

[20] We humans rarely notice how our brains store information. For example, most of us store our A, B, C's automatically in a singly linked list, not in a doubly linked list, and not in a random access data structure.

[21] One champion memorizer, Akira Haraguchi, learned 100,000 digits of **π** in 10 – 15 years. 15 years will suffice if the memorizer learns roughly 25 new digits per day, 5 days a week. Assuming 25 new digits can be memorized in 15 (concentrated) minutes, meaning 15 minutes from start to the first complete recital, and that each rehearsal of the 25 digits can be done in 1 minute, the memorization time for the first day in which 25 new digits are memorized would be 15 minutes for the new digits + 1 minute per rehearsal of these 25 digits at 15min, 30min, 60min = 1hour, 2 hours, 4 hours, 8 hours, 16 hours = 15+7 = 22min. To rehearse previously memorized digits on that same day would take 1 minute each for the digits memorized 1 day ago, 2days ago, 4dys, 8dys, 16dys, 1mo, 2mo, 4mo, 8mo, 1yr, 2yrs, 4yrs, 8yrs, 16yrs, 32yrs, 64yrs ago, which comes to 16min. The total is 22+16=38 minutes per day to achieve this awesome feat. Essentially anyone who cares enough for **π** to spend 15 years at it can do this. Haraguchi cares: he views pi as "the religion of his universe."

> ***COMM*** *(Communication Complexity) = length of (a precise description of) the humanly readable/humanly understandable algorithm PLUS length of the traces of execution on enough example(s) to cover all cases (execute every instruction and take every control flow arrow) in the algorithm.*

For example, suppose the preprocessing algorithm requires the user to memorize two randomly generated functions $f_1$, $f_2$, from the 26 letters to the 10 digits. To describe this operation, the algorithm must tell the human user exactly how to generate, memorize and compute these functions. Since memorizing two such functions can take twice as long as memorizing one, and since the two functions can be confused, a better algorithm might suggest to memorize a single function f from the 26 letters to the 2-digit numbers, 00 to 99, and then to set $f_1(x)$, $f_2(x)$ = most significant, least significant digits of f(x). These slightly longer more detailed instructions are crucial for keeping the preprocessing time under the stipulated 2 hours.

Because humans have different computation rates, we count steps, which are independent of the human, rather than running time. To relate step counts to running time, we give bounds on the time it typically takes a human to perform each step.

More precisely, we propose to use the pair < PREP, PROC> as the right measure of human effort to use a schema. This pair is typically a mix of numbers (for PREP) and formulas (a function of the challenge length, n, for PROC). We are not against using formulas for these counts; we are just against the typical emphasis on polynomial versus exponential step counts, as these distinctions are not useful for human computation.

## 3. Examples of Password Creation (and Re-Creation) Schemas

We present three schemas that illustrate the human computation model and associated complexity measures. For all three schemas, preprocessing requires generating and memorizing a single letter-to-digit map. The first two schemas are trivial; the third is not.

### 3.1. Letter substitution schema

**Preprocessing**. Memorize a single random (i.e. uniformly random) letter-to-digit map f. The memorization of a map from letters to digits can be done with 30 minutes of concentrated effort up front, and about the same amount of total additional time spent on spaced rehearsals (1 minute rehearsals at the 21 successive later times: 1, 2, 4, 8, 16 hours, 1 day, 2, 4 days, 1, 2 weeks, 1, 2, 4, 8 months, 1, 2, 4, 8, 16, 32, 64 years.

**Processing**. Given a challenge (string of letters) C, run the following algorithm:

1. Set a pointer to the first letter of C
2. Repeat until the entire challenge has been processed:
    2.1. Apply map f to current letter
    2.2. Output mapped value
    2.3. Shift pointer to next letter of challenge

Processing uses only 2 chunks of short-term memory: one pointer into the challenge and one for the mapped value of the current letter.

**Example:** Suppose the map from letters to digits is given by the position of the letter in the standard alphabetic ordering mod 10. Specifically A -> 1, B -> 2, … , I -> 9, J -> 0, K -> 1, … , Z -> 6. Then this letter substitution schema maps GMAIL -> 73192 and APPLE -> 16625.

**COMM** = (Description of preprocessing is less than 10 words; of processing is less than 40 words, and the description of the example is less than 60 words. Length of actual trace is 2 (initialization) + 3n steps on a challenge of length n) = 2 + 3n. Since n=5 for the two traces, total cost is less than 10+40+60+2(2+3×5) < 150.

**PREP** = 26 tosses of a 10-sided die to generate the random key plus 26 pairs to store it has a cost of 78. The lion's share of the cost is the hour it takes to memorize the key.

**PROC** = on a challenge of length n, this comes to: (n reads from long-term memory to compute the f function n times) + (2n reads and writes to short term memory) = 3n.

Modifications of this letter substitution schema might use special rules to avoid consecutive repeated letters. These include skipping a consecutive repetition or shifting up in the alphabet by 1 or 2 to avoid consecutive repetitions.  Example: AAA -> 123.

### 3.2. Single-digit Single-string schema

**Preprocessing**. Memorize a single random letter-to-digit map f as a hash function, and memorize a fixed random string S as a single chunk.

**Processing**. Given a challenge (string of letters), run the following algorithm:

1. Set current letter pointer to first letter of challenge
2. Set SUM = mapped value of current letter
3. Shift current letter pointer one step to the right in the challenge
4. Repeat till challenge is exhausted:
    4.1  Apply map to current letter
    4.2  Add newly mapped value to current SUM mod 10
    4.3  Shift current letter pointer to next letter of challenge
5. Output SUM  [Comment: Steps 1-5 generate and output just one digit]
6. Output the fixed string S.

Processing uses 3 chunks of short-term memory: one for a pointer into the challenge, one for the mapped value of the current letter and one for the running SUM.

**Example:** For illustration, let the memorized map be the letter position in the standard alphabet mod 10 (A -> 1, B -> 2, …, J -> 0, …, Z -> 6). Suppose the memorized fixed string is SESAME1@. Then for GMAIL we have G+M+A+I+L (mod 10) = 7+3+1+9+2 = 2, and so GMAIL -> 2SESAME1@.

**PROC** = [apply map + set SUM + shift pointer + (n-1) x (apply map + add mod 10 ) + output SUM + output S] = 3 + (n-1)(1+1.5)+ 1 + |S| = 2.5n + 1.5 + |S|.

The single-digit single-string schema produces only 1 digit that depends on the challenge, the rest of the response being a fixed string.  If a longer password depending on the challenge is desired, the user could produce more digits in a number of ways: (1) output a digit for the sum of every 3 letters (mod 10), or (2) make multiple passes over the challenge, using a different rule in each pass.  For example, in the first pass do as described above; in the second pass, output the alternating sum and difference of letters. So the second digit for GMAIL would be G – M +A-I+L = 7-3+1-9+2 = 8. As we will see in a subsequent section, there is a well-defined sense in which even the production of a single new digit based on the challenge achieves a well-defined quantified security.

### 3.3 Skip-To-My-Lou (STML) schema

The STML schema computes $F_x(C)$ as follows:

**Preprocessing**: Memorize a random map x as a hash function from letters to digits.

**Processing**:  Given a challenge C consisting of a string of letters, and the map x, do:

1. Set SUM_V = mapped value of last letter of challenge
2. Set POINTER (current letter) to first letter of challenge
3. Repeat until POINTER shifts past last letter of challenge:
    3.1. POINTER_V = mapped value of pointer
    3.2. Set SUM_V = (SUM_V + POINTER_V) mod 10
    3.3. If SUM_V is less than 5, output SUM_V.  [In any case, do not modify SUM_V]
    3.4. Shift POINTER from current letter in challenge by one letter to the right

**Output:** The string of SUM_V outputs produced by the above algorithm is $F_x(C)$.

**Example:**  Let the memorized map x be (A ➔ 1, B ➔ 2, …, I ➔ 9, J ➔ 0, …, Y ➔ 5, Z ➔ 6).  This maps the individual letters of GMAIL to 7, 3, 1, 9, 2.  The sequence of SUM_V values is 9, 2, 3, 2, 4, resulting in $F_x$(GMAIL) = 2324.

**PROC** = [apply map + set SUM + shift pointer + (n-1) x (apply map + add mod 10 + compare to 5 + output (maybe) + shift pointer)] = 3 + (n-1)(1+1.5+1+0.5+1) = 5n-2.

The output has expected length that is half the length of the challenge. To get a longer output, the user can append a fixed string to the challenge and run STML on the extended challenge. (Note that appending a fixed string to a challenge yields far better security than appending a fixed string to a password: if appended to the password, an adversary can determine the string from observation of two passwords, something she cannot do if the string is appended to the challenge.)

## 4. Humanly Computable One-Way Functions [22]

In complexity and cryptography theory, a One-Way Function (OWF) function F is defined to be any function (from strings to strings over some finite alphabet) such that (1) F(x) can be computed by a Turing machine on any input string x in time poly(|x|) [23], and (2) any Turing machine that runs on input y in poly(|y|) steps (for some fixed poly) has a negligibly small probability to invert F, i.e. to find a pre-image x' such that F(x') = y = F(x). One-way functions are conjectured to exist and play an important role in complexity theory and cryptography.

A *humanly usable*[24] *"one-way function"* (HU-"OWF") F is a function that (1) can be computed by a humanly usable algorithm[25], and (2) cannot be inverted on any but a "small" fraction of outputs y by a computer that has knowledge of the HU-"OWF" algorithm but not the key, and that executes a total of at most $10^{24}$ instructions.

Do humanly computable one-way functions exist? We propose a candidate here based on the STML schema: Let N be the size of a sufficiently large alphabet over which the map is defined, e.g. N=26 for Standard English alphabet and N=100 for 2-digit numbers. (N=10 will not do as it would permit an adversary to invert F by running through all possible x.) A random map x from N characters (letters and/or digits) to the 10 digits can be written as a string of N digits – the map values in a fixed ordering of the alphabet. Now generate and fix a **random** challenge C of sufficient length (roughly N ln N, the precise length to be determined) so that every character (typically letter or digit) is likely to be in the range. For a fixed such C, define the function $\mathbf{F}_C$ by $\mathbf{F}_C(x) = F_x(C)$, the function that takes as input the N-digit string x and outputs a string y by running STML with the map defined by x on a fixed challenge string C. Note that STML is being used here to compute $F_x(C) = y$ with x variable and C held constant, not x held constant and C variable.

The computational problem for the adversary - who knows C, $\mathbf{F}_C$ and y - is to find a string (map) x whose image under $\mathbf{F}_C$ is the observed y. We are not aware of any algorithm to solve this problem efficiently, i.e. to solve this inverse problem in less than $10^{24}$ steps, nor have we been able to show that an algorithm to invert the STML "OWF" would enable one to solve one of the hard problems that come up in machine learning (though we have considered only Gaussian elimination on equations that have errors and Gaussian elimination for sparse matrices).

Since the input x is now the key, not the challenge, a password schema that computes a HU-"OWF" has the important property that from knowledge of y, an adversary cannot determine the key, x.

If both x and C are unknown to the adversary, then not only is it hard for the adversary to determine x but also to determine C. This is no big deal if the adversary knows the challenge, which is what this theory assumes. But in practice, the user of a schema could modify challenges using some personal rule, such as the starting location (see the appendix). In that case, the adversary will not know the actual modified challenge either.

---

[22] Note that a humanly computable password generator is not in general either a "one-way function" or a "pseudo-random generator".
[23] |x| = length(x)
[24] The HU-"OWF" is *defined* to be humanly usable by the fact that a humanly usable algorithm can compute its output. It is *actually* humanly usable in the sense that humans can compute it, if the input is short enough (typically of length at most 25, but possibly more) and humans can perform the mandated operations in the mandated times (typically ⅓ to 1 second).
[25] This puts a strict upper bound on the length of input and/or time to compute the HU-"OWF". Furthermore, to be an interesting concept, the input length, n, must have a minimum size, say n > 24, which would ensure that the adversary cannot simply try all $10^{25}$ possible x's.

# 5. Humanly Computable "Pseudo-Random Generators"

A standard (cryptographically secure) **PRG** (Pseudo-Random Generator) is an algorithm based on a parameter/key **k** and a polynomial poly, poly(n) > n, that maps a (random) input string, the challenge/seed, of length **n** to a random-looking output string of length poly(**n**), the pseudo-random string. That output "looks" random in the sense that any poly-time randomizing Turing machine that has knowledge of the general algorithm for computing the PRG, but not the key, k, cannot distinguish a string output by the PRG from a truly random string R of the same length with negligible probability of error.

A **"PRG"** is an algorithm having a parameter/key **k** that maps a (random) input string, the challenge/seed, of length **n** to a random-looking output string of length **2n**, the "pseudo-random" sequence.[26] That output "looks" random in the sense that a computer that has knowledge of the "PRG" algorithm (but not the key, k) and that executes a total of at most $10^e$ instructions for some specified e (which may be as large as e = 24 or as small as e=12) cannot distinguish[27] a string output by the "PRG" from a truly random string R of the same length with probability of error < 1/4.

The "PRG" is a HU-"PRG" if its output can be computed by a humanly usable algorithm *and therefore, assuming the seed is short enough, by humans that can perform the mandated operations in the mandated times (typically 1 minute)*.

We propose a quasi humanly computable HU-"PRG" based on STML. It is not a true HU-PRG for several reasons, the most important of which is that we have no proof (not even one based on reasonable assumptions) that the output is "pseudo-random," we have only a hope, which we have not been able to translate into a proof.

Other reasons that the HU-"PRG" is not a true HU-PRG:

1. Fewer than 2n digits may be generated. (There are several ways to skirt this, such as by increasing n, the size of the challenge.)
2. More than 2n digits may be generated (while keeping track of how many digits have been produced puts an additional burden on already heavily burdened human, she could mark the end of 2n output spaces and stop writing the output when the marker is reached).
3. More generally, the need for more pointers than are available: we can manage to keep track of skip-length on the fingers of one hand, and to track the length of the output on the fingers of the other hand. Such extraordinary actions are not permitted in this model (for doing computations entirely in one's head) - unless and except when it is explicitly pointed out that they are being used.

Our HU-"PRG" takes as input an n-long string of digits in the set {0,1,2,3,4} and outputs a random-looking string of approximately 2n digits in the same set, {0,1,2,3,4}.

The intermediate work of the algorithm involves all ten digits {0,1, …, 9}. That work is private (except for whatever it outputs whenever it outputs it), and except as noted in 3 above can be performed in the user's tiny STM with the few allowable pointers plus digits from the set {0,1, …, 9}.

The HU-"PRG" uses STML as a subroutine. When used as a password schema, STML uses a letter-to-digit map; for PRG, it uses a digit-to-digit map.

    The basic idea:
        Initialize:
            1. Set SUM = last digit of the input
            2. Set pointer to first digit of input
            3. Until pointer drops off the challenge:
                3.1. Set SUM = mapped SUM [comment: the map is from digits to digits]
                3.2. Set SUM = SUM + current digit (mod 10)
                3.3. If SUM is less than 5, output SUM
                3.4. Shift pointer to next digit of input

In detail, with an example:

Preprocessing: Memorize a single digit-to-digit map.  EXAMPLE: map is T(i) = i+1 (mod 10)

---

[26] The output of a PRG on a seed of length k is typically only required to be of length k+1, which can be extended to longer lengths by reapplication of the PRG. Such reapplications make the PRG humanly unusable for many reasons, including that the human has no place to store the intermediate k+1 digits, which are needed as seed to generate the next k+2 digits.

[27] "Indistinguishable" is usually up to a negligibly small 1/poly(n) factor. In human usability, this must be replaced by a concrete epsilon.

Processing: Given a challenge C of length n:        EXAMPLE: C = 3141592653
1. Set SUM = last digit of C                                            SUM = 3
2. Set pointer to first digit of C.                                        pointer = 3
3. Repeat till end of challenge:
   3.1 SUM = map (SUM)
   3.2 SUM = SUM + digit of challenge given by the pointer (mod 10)
   3.3 If SUM ≤ 5, output SUM
   3.4 If not End-of-Challenge, shift pointer to next digit of challenge

Iter 1: T(3) = 4, 4+3 = 7, no output
Iter 2: T(7) = 8, 8+1 = 9, no output
Iter 3: T(9) = 0, 0+4 = 4, output 4
Iter 4: T(4) = 5, 5+1 = 6, no output
Iter 5: T(6) = 7, 7+5 = 2, output 2
Iter 6: T(2) = 3, 3+9 = 2, output 2
Iter 7: T(2) = 3, 3+2 = 5, no output
Iter 8: T(5) = 6, 6+6 = 2, output 2
Iter 9: T(2) = 3, 3+5 = 8, no output
Iter 10: T(8) = 9, 9+3 = 2, output 2

3141592653 -> 42222

Memory requirements: 1 digit (SUM), one pointer into challenge.

A password schema that is also a pseudo-random generator has some useful properties compared, say, to the schema that replaces each letter of the challenge by its hash value under the key. For one, it can make it more difficult for the adversary to get a handle on the user's key.

The HU-"PRG" that we propose is based on a key, k, which is a string randomly chosen from the $10^{10}$ strings of digits of length 10. The string k = $k_1 k_2 \ldots k_{10}$ represents the hash function 1 -> $k_1$, 2 -> $k_2$, … 0 -> $k_{10}$. The schema takes as input a random seed/string of digits C of a length n (to be determined), whose digits have been randomly chosen from the set {0,1,2,3,4}. The initial output $F_k(C)$ of the PRG is the output of STML applied with key k to challenge C. This string $F_k(C)$ is rarely long enough to be the complete output of the HU-"PRG," as its expected length is just half the length of C. To produce a longer string, at the cost of one more (expensive) pointer, STML is run on many substrings of C and the results concatenated. The substrings of C have to be chosen in a humanly usable way. For a seed C of length n, let $C_i$, be the substring of characters obtained by skipping the first i characters of C, then including the next i characters, skipping the next i, including the next i, and so on. E.g., $C_1$ is the substring of characters in even positions, and $C_2$ is the substring of characters obtained by alternately skipping two and including two. To be humanly usable (if only barely), keep track of the skip length on one's fingers.

**STML HU-"PRG" 1:** For a seed C of length n, the output of the **HU-"PRG" 1** is the concatenation $F_x(C) \cdot F_x(C_1) \cdot F_x(C_2) \cdot \ldots \cdot F_x(C_{(n-1)/2})$. In the first application of STML, the carry digit is the default last digit of the challenge. From the second application onwards, the carry digit is the running sum mod 10 from the previous iteration (regardless of whether the digit was output).

The expected length of the output is $\frac{n + \frac{n}{2} + \cdots + \frac{n}{2}}{2} = \frac{n\left(1 + \frac{n-1}{2}\right)}{2} = \frac{n(n+1)}{4}$.

Example: Suppose the seed is 31415926 [28] and the key is k(i) = 3i mod 10. Then starting with the last digit, 6, as carry and using the entire challenge as the seed, this maps in the first round to 14692577, which maps to 142. Then for the substring of the challenge consisting of digits in even positions (1196), with carry 7 from the previous round, we get 2706 and the output is 20. Next by skipping two and using two, we have the substring 4126. With carry 6, this maps to 2735 and the output is 23. Then skipping 3 digits and using 3 digits, we have the substring 159. With carry 5, this maps to 638, so the output is 3. Next is the substring 5926, which, with carry 8 maps to 9606 and the output is 0. Thus the output is 142202330.

The following theorem proves that the STML HU-"PRG"1 can be broken for challenges of length n = 10. The argument fails for n = 20.

**Theorem:** STML HU-"PRG"1 can, with high probability, be broken for challenges of length n = 10 using at most $10^{15}$ operations.[29] (At $10^{11}$ operations per second, this takes at most 3 hours on a laptop.)

---

[28] For symmetry, we might (but don't) require that the seed digits be randomly chosen from {0, 1, … 5}.
[29] This is an upper bound. Is there a more efficient way? One might conjecture that no method can break the "PRG" in less than $10^n = 10^{10}$ steps.

**Proof:** On a challenge of length n, we require that the output be of length 2n. These 2n digits of output must come from half of approximately 4n locations. Guess the 2n of 4n locations from which the output comes. There are $2^{4n}$ possible choices ($2^{40} \approx 10^{12} = 1000 \times 10^9$). For each choice, we get a set of 2n linear equations mod 10 in 10 variables (one for each digit of the challenge, after applying the key). Of these, there will likely be n linearly independent ones. They can be solved in $10^3 = 1000$ operations. This gives a total of $10^{15}$ operations[30,31] ∎

We now define a second, simpler candidate. It uses a single digit-to-digit map as before and produces an output of length twice the length of the challenge *in expectation*[32]. The function, $F_x(\cdot)$, and substrings $C_1, C_2$ of the challenge are defined as above in STML HU-"PRG"1.

**STML HU-"PRG"2:** On a challenge C of length n:

$$\text{Output} = C \cdot F_x(C) \cdot F_x(C_1) \cdot F_x(C_2).$$

This HU-"PRG"2 proceeds exactly like STML HU-"PRG"1, but only for 4 passes. In 4 passes a string of length n is expected to result in an output of length n+(n/2 + n/4 + n/4) = 2n.

As an example, suppose the challenge (seed) is 31410421 and the key is k(i)=3i mod 10. Then in the first pass, with the last digit 1 as carry, the challenge maps to 691420443 and the output is 1420443. The second pass the subchallenge 1141 with carry 3 maps to 0172 and the output is 012. The third pass, with carry 2, maps the subchallenge 4121 to 0156 with output 01. Thus the entire output is 314104211 1220443 012 01.

It is a tantalizing possibility to generate a longer string by allowing the human to use previously generated output in order to generate more output.

We note that the seed could be a letter string and the key a letter-to-digit map, in place of a digit string and a digit-to-digit map respectively as above. When the letter string has additional meaning, e.g., it is a word or phrase, then the human might not even need to have the seed in writing while running **HU-"PRG"1** and only need **to hear or see it once.**

## 6. Conclusion

We have presented a precise model for human computability, focusing on the simple setting in which all processing is done in the human head, with no pencil, paper or any other kind of external device besides what is necessary to present the input. The model allows us to quantify the complexity of humanly usable algorithms (schemas).[33]

## References

Blocki, J. (2014). Usable Human Authentication: A Quantitative Treatment. Ph.D. Thesis. *CMU-CS-14-108.pdf*

Blum, M. & Micali, S. (1982). How to generate cryptographically strong sequences of pseudo-random bits 1982. *SIAM Journal on Computing*, (13): 850–864, 1984. Preliminary version in 23rd FOCS, 1982.

---

[30] Compare to the $10^{12}$ mentioned in footnote 12.
[31] At $10^{11}$ operations per second, it takes less than 3 hours to find a challenge that works… and at least one definitely works.
[32] Because the output has expected as opposed to guaranteed length twice that of the challenge, STML HU-"PRG"2 is NOT a true "PRG".
[33] Natural extensions of this model include:
 (1) External Pointer: Again, no paper or pencil allowed, but the user can read the input challenge during processing using a finger as pointer into the challenge. An example of this is our candidate HU-"OWF". For inputs that are random (or unfamiliar) long digit sequences, the human will have to read the input as she applies the HU-"OWF".
 (2) Write-once Tape: A pencil is allowed but no eraser, and the user is allowed to read only what she has already written. Games such as Sudoku and Crossword puzzles involve writing the output. During this process, previously written output is crucial to determining the remaining output.
 (3) Write-once Tape and External Pointer: The use of a pencil as in (2) above, and a finger as a pointer are allowed. As an example, we can extend our second candidate for a Humanly Usable PRG, namely STML HU-"PRG"2, to produce a longer output (of expected length roughly quadratic in the input length): We run STML HU-"PRG"2 as defined. The output includes the challenge and in expectation, n more digits. We then run the entire schema again, on the additional digits after the challenge, treating them as the new challenge. This process can be repeated till no further output is produced. The user has to be able to read their own output so far. One way to make this easier is for the user to write the output of each pass on a new line.

Blum, M., & Vempala, S. (2015). Publishable Humanly-Usable Secure Password Creation Schemas. *HCOMP2015*. 32-41,

Bonneau, J. (2012). The science of guessing: analyzing an anonymized corpus of 70 million passwords. *IEEE Symposium on Security and Privacy (SP),* 538–552.

Goldreich, O., (2004). Foundations of Cryptography – A Primer. *Foundations and Trends in Theoretical Computer Science*, NOW publishers, 1(1): 1-116.

Ives, B., Walsh, K. R., and Schneider, H. (2004). The Domino Effect of Password Reuse. *Communications of the ACM,* April 2004, 47(4): 75-78.

Li, Z., He, W., Akhawe, D. and Song, D. (2014). The Emperor's New Password Manager: Security Analysis of Web-based Password Managers. *USENIX Security,* 465-479.

Kruger, H., Steyn, T., Medlin, B. and Drevin, L. (2008). An empirical assessment of factors impeding effective password management. *Journal of Information Privacy and Security*, 4(4):45-59.

Pimsleur, P. (1967). A Memory Schedule. *The Modern Language Journal*, 51(2): 73-75.

Shay, R., Komanduri, S., Durity, A. L., Huh, P. (S.), Mazurek, M. L., Segreti, S. M., Ur, B., Bauer, L., Christin, N. and Cranor, L. F. (2014). Can long passwords be secure and usable? *CHI 2014:* 2927-2936.

Woźniak, P. A., & Gorzelańczyk, E. J. (1994). Optimization of repetition spacing in the practice of learning. *Acta Neurobiologiae Experimentalis*, 54: 59-62.

Yao, A. (1982). Theory and application of trapdoor functions. *23rd IEEE Symposium on Foundations of Computer Science*, 80–91.

# APPENDIX:  Creating Personal Private Password Schemas

This paper defines password schemas, gives a foundation for their development, and provides some guarantees for them.  For users who opt to create their own schemas – for the security that comes of keeping both schemas and data (not just data) private - we present here some ideas for creating and analyzing private schemas.

## A sampling of methods for generating passwords:

**Typewriter-shift**: type the password by shifting the position of the fingers on the keyboard left ←, right→, up↗, up↖, down↘, down↙ or some combination of these.  For example, shifting right by a single key maps **password** to **[sddeptf**.  A more complex shift might be a kind of knight's move →↗ that maps **password** to **=err4-6t**.  An even more complex rule would be one that replaces →↗ by →↘ if and when this is necessary to ensure that no letter appears twice in a row:  **password** to **=erc4-6t**.

**Telephone number encoding**: define shift-addition by (for example) X+1=Y (the letter immediately following X in the alphabet), X+2=Z, X+3=A, and so on.  Take the first 3 or so letters of the challenge and shift each letter by the digits of a random "area code."  Choose digits at random from {1, 2, 3, 4, 5}.  For example, area code 412 shifts HELP -> <H+4, E+1, L+2, P> = LFNP.  More generally, use a random "telephone number," for usability made up entirely of digits chosen from {1, 2, 3, 4, 5}.  For example, telephone number 314 154-2153 maps AAA -> DBEBFECBFD.  This is humanly usable.   A more humanly usable number – at the expense of some security - is 314 111-1111.  Beware using 111 111-1111, however: a Google search for NBHJD, a string obtained in this way, reveals that NBHJD appears in MVFODZ 5IF NBHJD MJOL (luency 4he magic link).

**Start at an unexpected location in** the challenge:
Instead of starting at the first letter of the challenge:
        Start at the last letter.
        Start at the 2nd vowel.
        Start 2 past the 1st vowel.  For example, on challenge AMEX, start at E; on challenge BANK, start at
        the letter K.
        Start 1 past the first letter of one of the following types:
- Letters that when capitalized contain a vertical line segment.  Specifically, these are the letters B, D, E, F, H, I, J, K, L, M, N, P, R, T, U, Y; if no such letter is in the challenge, then start at its last letter.
- The letter sounds like eee, namely one of B, C, D, E, G, P, T, V, or Z, if such a letter appears in the challenge; else the last letter.
- Vocalizing the letter requires the tongue to touch the upper palate, namely D L N T W.

**To handle repeats**: simply delete them. Or increment each letter by the number of times that that particular letter has (already) been seen. This may introduce repeats, but no matter: the purpose here is to make it harder for the adversary to zero in on what is going on.

**Carries**: for integer challenges, start with a carry that depends in some deterministic way on the challenge, perhaps setting the carry to the sum of digits mod 10 (or more generally to the dot product of the challenge (viewed as a vector) and a fixed random vector mod 10).

**Append a fixed string**: first generate a preliminary password as suggested above, then attach a fixed string such as aA1@, placing the fixed string at the beginning, end or interspersed with the letters of the password in some fixed way. Attaching the same four symbols in the same way to every password does not add much in the way of security but it helps to satisfy password requirements, and if the same string of four symbols is attached in the same way to all passwords, then this string will be self-rehearsing.

To you brave folks who create your own schemas, we recommend that you test your schemas with at least two password checkers. Don't test actual passwords. Just test the proposed schema's responses (passwords) to a sample of randomly generated challenges.

We recommend two checkers, a password meter such as www.passwordmeter.com and www.google.com

**Google as checker**: if a hacker has a list of passwords to check, then Google is likely to recognize all the words in that list. By "Google does not recognize **fun924aA1@**," we mean that Google responds to the query **fun924aA1@** with something like: "Your search – **fun924aA1@** – did not match any documents."

**Password meter**: in our experience, if we end every password with the same four symbols, e.g. **aA1@** then passwordmeter.com is much happier than without those four symbols. In 2016, for example, passwordmeter.com gave **fun924** a rating of 38% and **fun924aA1@** a 100%.